\documentclass[journal]{IEEEtran}

\usepackage{xcolor,soul,framed} 

\colorlet{shadecolor}{yellow}
\usepackage[pdftex]{graphicx}
\graphicspath{{../pdf/}{../jpeg/}}
\DeclareGraphicsExtensions{.pdf,.jpeg,.png}

\usepackage[cmex10]{amsmath}
\usepackage{array}
\usepackage{mdwmath}
\usepackage{mdwtab}
\usepackage{eqparbox}
\usepackage{url}

\begin{document}
\title{A Novel Field-Free SOT Magnetic Tunnel Junction With Local VCMA-Induced Switching}
\author{Rui Zhou, Haiyang Zhang, Hao Wang, \IEEEmembership{Member, IEEE}, Jin He, \IEEEmembership{Senior Member, IEEE}, Qijun Huang, and Sheng Chang, \IEEEmembership{Senior Member, IEEE}
\thanks{ }
\thanks{ }
\thanks{The authors are with School of Physics and Technology, Wuhan
University, Wuhan 430072, China (e-mail: wanghao@whu.edu.cn).}
\thanks{S. Chang is also with School of Microelectronics, Wuhan University, Wuhan 430072, China (changsheng@whu.edu.cn).}}

\maketitle

\begin{abstract}
By integrating the local voltage-controlled magnetic anisotropy (VCMA) effect, Dzyaloshinskii-Moriya interaction (DMI) effect, and spin-orbit torque (SOT) effect, we propose a novel device structure for field-free magnetic tunnel junction (MTJ).  Micromagnetic simulation shows that the device utilizes the chiral symmetry breaking caused by the DMI effect to induce a non-collinear spin texture under the influence of SOT current. This, combined with the perpendicular magnetic anisotropy (PMA) gradient generated by the local VCMA effect, enables deterministic switching of the MTJ state without an external field. The impact of variations in DMI strength and PMA gradient on the magnetization dynamics is analyzed.
\end{abstract}


\section{Introduction}
\label{sec:introduction}
\IEEEPARstart{M}{agetic} Random Access Memory (MRAM) has gained prominence in the field of memory storage due to its non-volatility, low power consumption, and high compatibility with CMOS technology\cite{10264021,9270597,2019,8778100}.
Currently, the main storage element in MRAM is the magnetic tunnel junction (MTJ)\cite{9235579} with perpendicular magnetic anisotropy (PMA).
 Achieving deterministic switching of the free layer magnetization in the MTJ without an external magnetic field is crucial for further development of MRAM.
 In recent years, many explorations have been made to achieve deterministic field-free spin orbit torque (SOT) \cite{baek2018spin} switching for vertical magnetization.
 Some approaches involve the use of induced spin textures, such as domain wall structures\cite{9072283,10106129} or skyrmion structures\cite{LUO2020165739,Tokura2020,7482654}, to achieve field-free information transfer under specific conditions.
 Another class of promising methods involves using exchange bias to induce an internal equivalent magnetic field for field-free transmission\cite{8993513,10.1063/5.0156241}.

Recent studies have shown that the chiral symmetry breaking induced by the Dzyaloshinskii-Moriya interaction (DMI) effect\cite{moriya1960anisotropic} can enable stable field-free switching of the MTJ\cite{yu2014switching}.
 However, in the practical large wafer-scale fabrication process, achieving reproducible wedge layers\cite{wu2020chiral}, poses a challenging task that requires further exploration.
 The local voltage-controlled magnetic anisotropy (VCMA) effect\cite{amiri2015electric} that has been widely used to control the movement of magnetic skyrmions\cite{kang2016voltage,lone2022skyrmion}.
 Therefore, it's worth trying to apply the local VCMA effect to MTJs with DMI to achieve deterministic switching without an external field.

In this letter, we conducted an analysis of uniformly magnetized ferromagnetic materials with isotropic gradient under the influence of SOT using the micromagnetic simulation framework $Mumax3$\cite{vansteenkiste2014design}.     On this basis, we propose a novel asymmetric MTJ structure LVA-MTJ (local voltage controlled anisotropic magnetic tunnel junction), which can control MTJ by combining local VCMA effect and DMI effect. The stability of the device is analyzed by considering the range of existence of uniform magnetic states in small size ferromagnetic elements, and the influence of DMI strength and magnetic anisotropy gradient on the MTJ switching behavior is investigated.

\section{Theory And Simulation}

\begin{figure}[h]
\centerline{\includegraphics[width=\columnwidth]{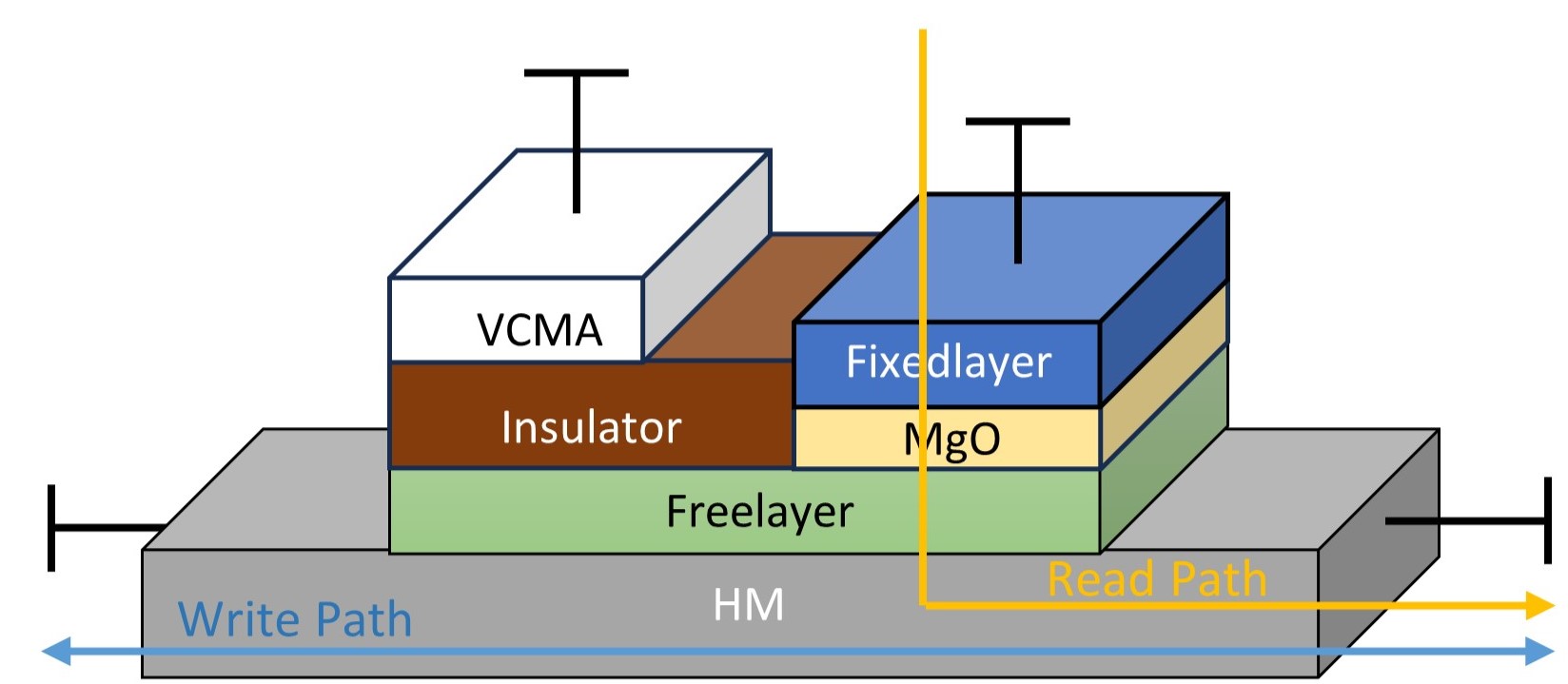}}
\caption{Illustration of the LVA-MTJ. }
\label{fig1}

\end{figure}

Fig.\ref{fig1} illustrates the schematic diagram of the proposed LVA-MTJ.
 Compared with the traditional SOT MTJ, this device introduces a heavy metal layer with a strong DMI effect, so that the free layer can produce chiral spin texture when the SOT current is applied.
 Furthermore, the free layer is extended, and an insulator layer is added in the extended part to realize the local VCMA effect, which is used to produce PMA gradient. 
 The combination of these effects enables field-free switching of the MTJ.

In this device, the relative strength between exchange interaction and DMI interaction is critical.
 The exchange interaction\cite{aharoni1997introduction} is an inherent interaction in ferromagnetic materials, and its equivalent magnetic induction strength is expressed as:
\begin{equation}\hat{H}_{ex} = -2\sum_{i,j}A_{ij}{m_i}\cdot{m_j}\label{eq3}\end{equation}
where $A_{ij}$ is the exchange interaction strength between \textbf{${m_i}$} and \textbf{${m_j}$}, which causes adjacent magnetic moments to align parallel to each other.
 The DMI effect\cite{dzyaloshinsky1958thermodynamic} is induced by the bottom heavy metal channel acting on the MTJ free layer, and its equivalent magnetic induction strength is expressed as:
\begin{equation}\hat{H}_{dmi} = -\sum_{i,j}D_{ij}{m_i}\times{m_j}\label{eq2}\end{equation}
where $D_{ij}$ is the DMI tensor between ${m_i}$ and ${m_j}$, which causes adjacent magnetic moments to align perpendicular to each other.

\begin{figure}[b]
\centerline{\includegraphics[width=\columnwidth]{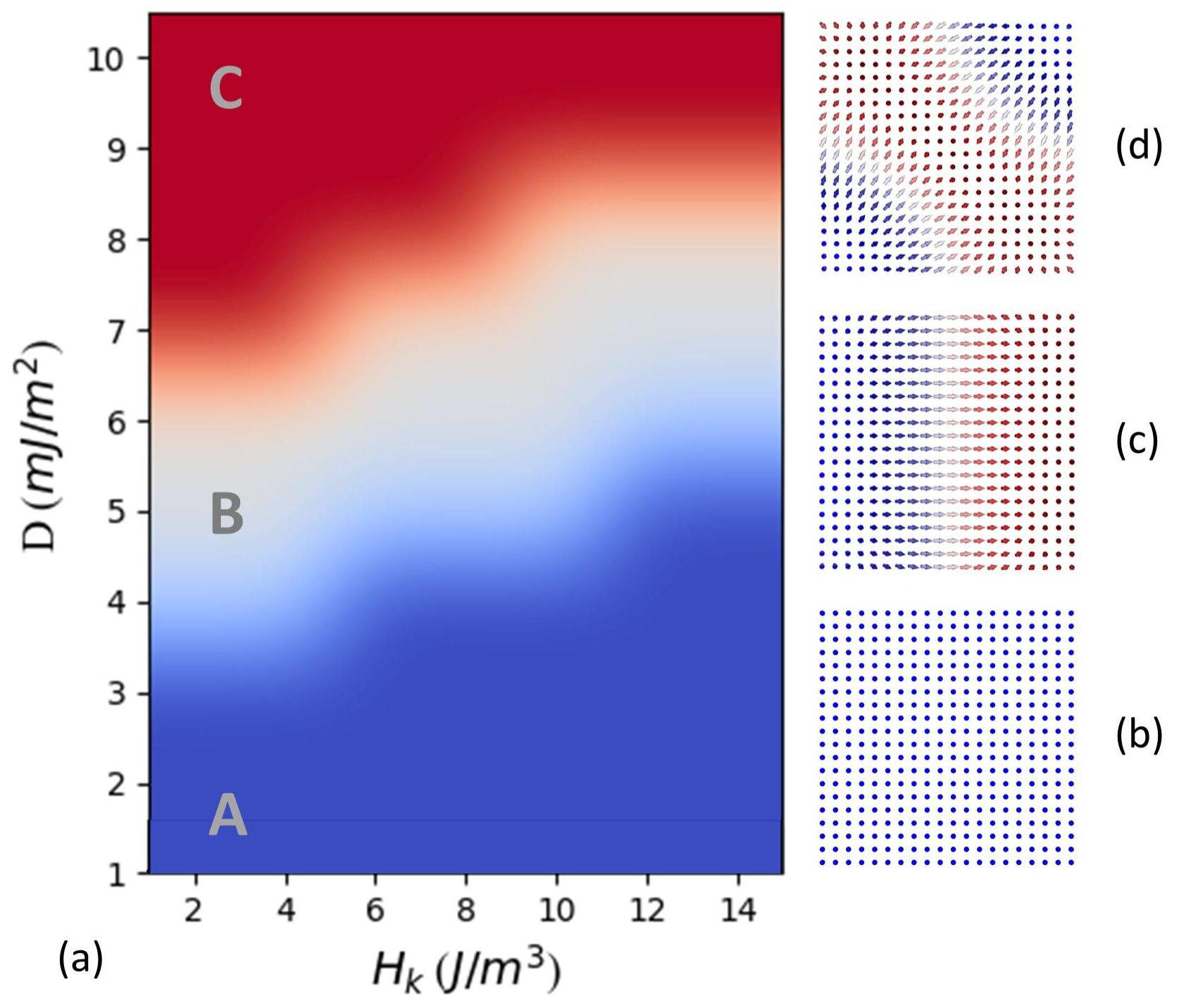}}
\caption{(a) Diagram of magnetic moment equilibrium state of the free layer under different magnetic parameters, in which the blue, white and red parts correspond to (b) uniform state,(c) vertical fringe structure and (d) circular fringe structure respectively.}
\label{fig2}
\end{figure}

With the PMA strength held constant, the curling of the free layer magnetization increases gradually with the enhancement of DMI strength.
 In this study, micromagnetic simulations are performed to investigate the magnetization states of a $20nm\times20nm\times1nm$ free layer under varying perpendicular magnetic anisotropy and DMI strengths at equilibrium. 
 The simulation results are shown in Fig.\ref{fig2}.

 As shown in Fig.\ref{fig2} (b), when the relative strength of the PMA is significant, the magnetization of the free layer is in region A, exhibiting a uniform state.
 The vertical stripe structure observed in region B (Fig.\ref{fig2} (c)) characterizes the relative strength of the DMI effect being greater than the PMA at this point. 
 Further curling of the magnetization into circular striped structures (Fig.\ref{fig2} (d)) is observed in region C.
 In order to ensure that the free-layer magnetic moment can be effectively read through the MTJ, it is necessary to ensure that the MTJ remains uniformly magnetized in the absence of SOT currents (that is the region A in Fig.\ref{fig2} (a)).

Referring to the experimental results of DMI effect\cite{yuasa2013future} and widely used micromagnetic simulation parameters\cite{buttner2017field,tretiakov2016creep}, the parameters of the free layer of LVA-MTJ device are as follows: the device size is $20nm\times20nm\times1nm$, the cell size in $mumax3$ is $1nm\times1nm\times1nm$.
 The saturation magnetization strength ($M_s$) is set to ${1.1\times 10^{6}} A/m$, the exchange strength ($A_{ex}$) is ${1.6\times 10^{-11}} J/m$, the uniaxial perpendicular magnetic anisotropy strength is ${8\times 10^{5}} J/m^3$, the damping factor ($\alpha$) is $0.1$, the DMI strength (Dind) is ${1\times 10^{-3}} J/m^2$, the SOT strength coefficient (SOTxi) is $-2$, and the polarization strength is $0.15$.

\begin{figure}[h]
\centerline{\includegraphics[width=\columnwidth]{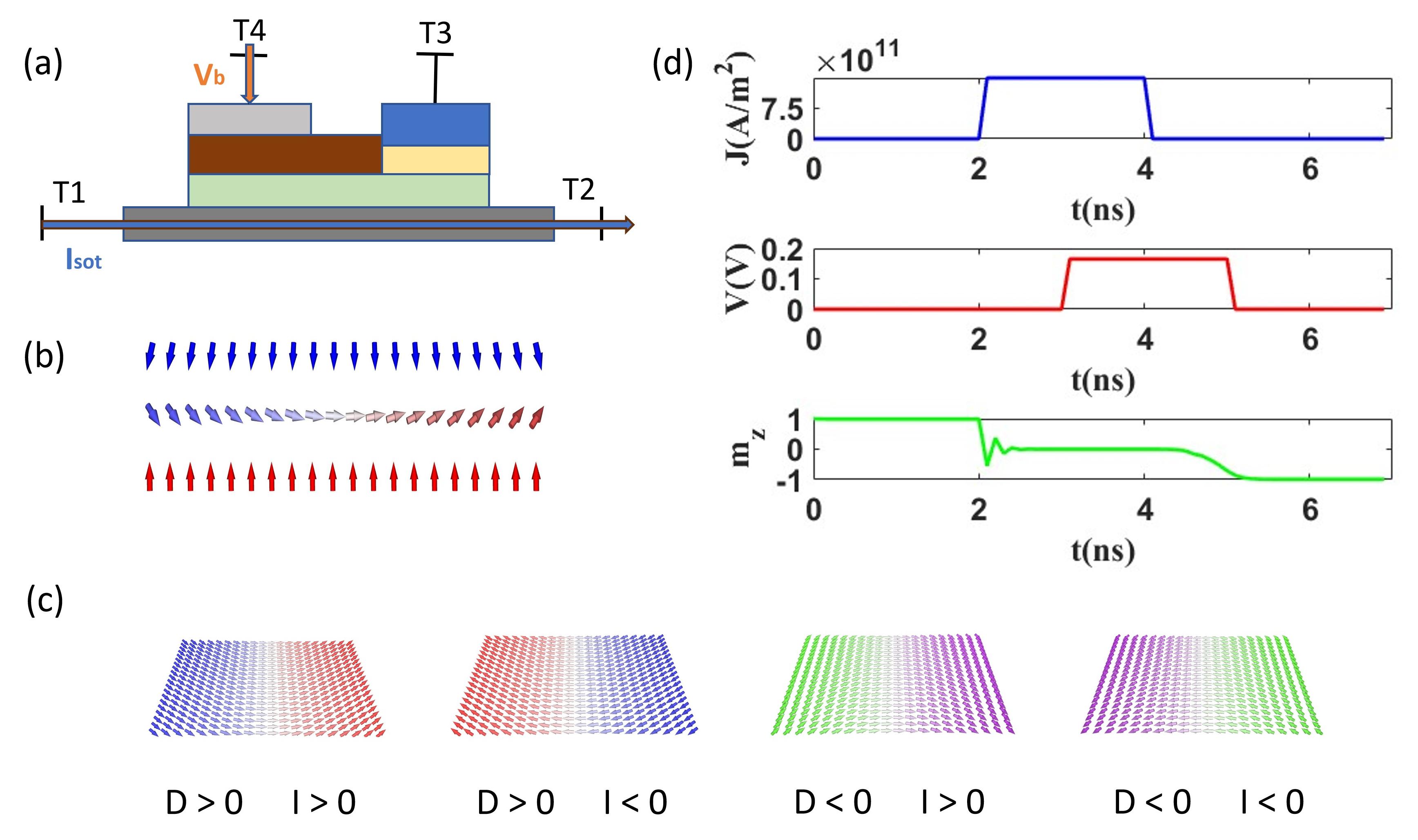}}
\caption{(a) LVA-MTJ plane schematic diagram.
(b) Free layer magnetic moment under SOT current and VCMA voltage.
(c) Spin textures under different DMI intensity and current polarity.
(d) The magnetic moment of the free layer varies with voltage
and current during the writing process.}
\label{fig3}
\end{figure}

The planar schematic of LVA-MTJ, as shown in Fig.\ref{fig3} (a), features four input and output ports. 
 Among these, T1 and T2 serve as the current paths of SOT, influencing the free layer magnetization through the Spin Hall Effect (SHE) effect\cite{sinova2015spin}. 
The ports T3 and T2 constitute the MTJ structure and are used for the readout of the free layer magnetization. 
 T4 is the VCMA voltage port, which adjusts the local PMA strength by altering the magnitude and polarity of the voltage to influence the flipping of the free layer magnetization.

Before elaborating on the working steps of LVA-MTJ, we need to understand the variation of free layer magnetization of the device under different voltage and current conditions.

 As shown in Fig.\ref{fig3} (b), when no SOT current and VCMA voltage are applied, the free layer magnetization aligns uniformly (either up or down spin) under the combined influence of PMA and exchange interaction.
 When SOT current is applied, the in-plane spin current generated by SHE effect weakens the effect of PMA. Under the influence of DMI effect, the magnetic moment of the free layer presents a chiral spin texture, and the magnetic moment of the free layer is determined by the direction of SOT current and the sign of DMI tensor, as shown in Fig.\ref{fig3} (c).
 At this time, the application of bias voltage produces a magnetic anisotropy gradient, which makes the free layer magnetic moments in different regions subject to different PMA (The positive bias voltage is used as an example).
 However, since the DMI effect is dominant, the chiral spin texture does not change.
Maintaining the  voltage and ceasing the SOT current, the free layer magnetization will return to a uniform state under the influence of exchange interaction.
 In this scenario, regions with stronger anisotropy exhibit greater flipping force, thus the final magnetization of the free layer will uniformly align towards the direction of stronger anisotropy.
 Therefore, deterministic reversal of the free layer magnetic moment can be achieved without an external magnetic field, enabling information writing.

From Fig.\ref{fig3} (d), it can be observed that the write process of LVA-MTJ consists of three main steps.
 The first step involves applying a SOT charge current of $150 MA/cm^2$ to the free layer from the T1 and T2 ports to induce the formation of a chiral spin texture.
 In the second step, the SOT current is maintained and a voltage of $\pm0.165V$ is applied to terminal T4 to modify the magnetic anisotropy.
 The functional relationship between the voltage and the anisotropy is determined as follows\cite{lee2017analog,kang2017modeling}:
\begin{equation}H_k(V_{b}) = H_k(0)-{\beta V_{b}}/t_{ox}\label{eq1}\end{equation}
where $\beta$ is the VCMA coefficient, here is ${9.0429\times 10^{-5}} J/V$ , and $t_{ox}$ is the thickness of the insulator layer.
 In the third step, the SOT current is removed while keeping the voltage at the T4 terminal unchanged.
 Under the influence of the differences in magnetic anisotropy, the magnetization undergoes directed flipping, completing the information writing process. 
 The resistance state of the MTJ can be read directly by applying current to the T3 and T2 terminals.

\begin{figure}[h]
\centerline{\includegraphics[width=\columnwidth]{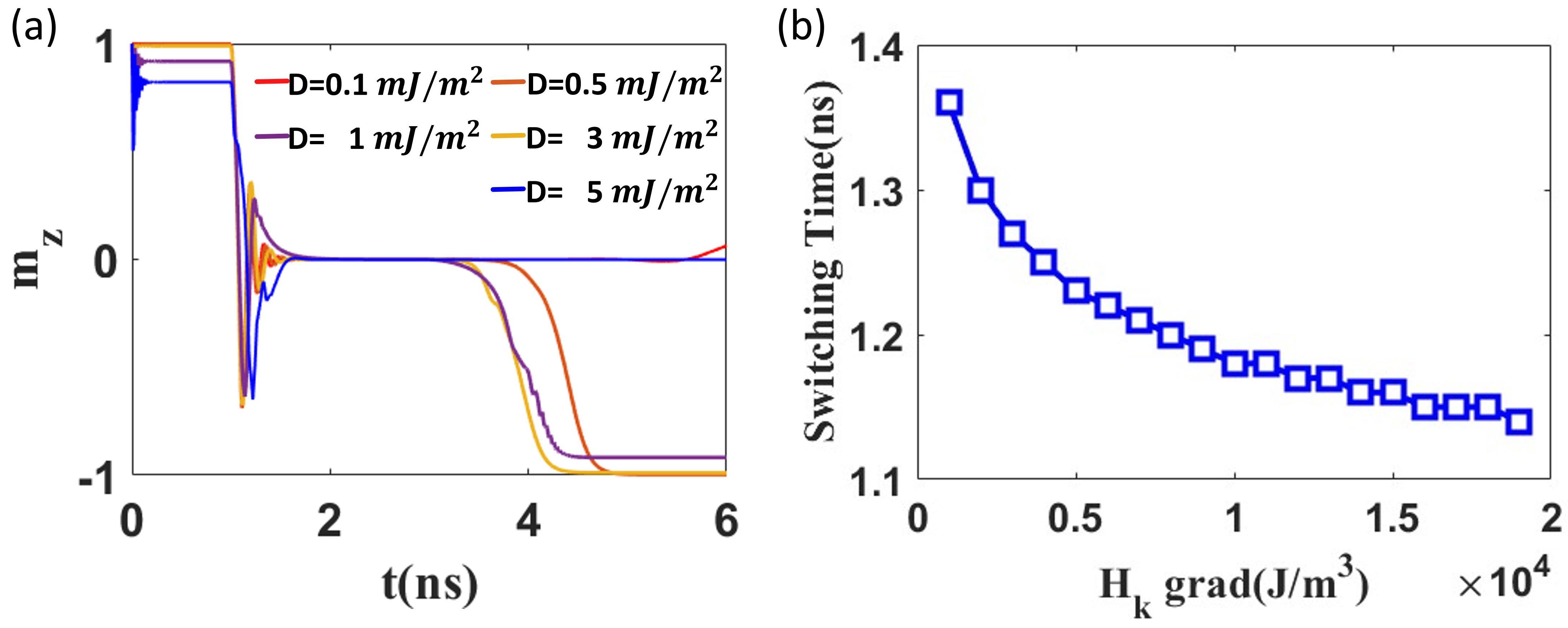}}
\caption{(a)The process of information inscribing varies with DMI strength.(b)The change of the time for the free layer to return to the uniform state with the anisotropy gradient after the release of SOT current.}
\label{fig4}
\end{figure}

The magnitude of the anisotropy gradient and variations in DMI strength have an impact on the information writing process of the device.
 On the one hand, as shown in Fig.\ref{fig4} (a), the effect of DMI strength on the information writing process is complex.
 When the DMI strength is too weak, an effective chiral spin texture cannot be formed to support the directed flipping of the free layer.  After applying the SOT current, the free layer magnetization remains in an intermediate state for a relatively long time.
 Conversely, when the DMI strength is too strong, even the chiral spin texture formed by removing the SOT current is difficult to break by the perpendicular magnetic anisotropy field, ultimately remaining in the intermediate state.
 In the intermediate state between the two extremes, an increase in the DMI effect leads to a gradual strengthening of magnetization fluctuation during the formation of the chiral spin texture under the influence of the SOT current.
 Furthermore, the recovery time after removing the SOT current also gradually lengthens.
 This reflects the weakening effect of DMI on the uniform magnetic state.

On the other hand, the variation in the magnitude of magnetic anisotropy gradient has little effect on the formation of self-selected textures when subject to SOT current.
 However, it slightly alters the time required for the free layer magnetization to recover from a chiral spin texture to a uniform magnetic state during the release of the SOT current, as depicted in Fig.\ref{fig4} (b).
 More specifically, the increase in the average magnetic anisotropy gradient  (decrease in the average value of the anisotropy field) leads to a reduction in the time required to return to a uniform magnetic state.

\section{Conclusion}
In this work, a field-free MTJ device based on the localized VCMA effect and DMI effect-mediated SOT switch is proposed.
 By locally applying the VCMA effect to the free layer to construct a PMA gradient, and ultimately combining it with the non-collinear spin texture formed by the DMI effect under the SOT current, deterministic switching of the free layer magnetization in the MTJ can be achieved without an external magnetic field.
 Through micromagnetic simulations, we effectively analyze the stable states and operating modes of the LVA-MTJ free layer. In addition, the scalability of DMI intensity and PMA gradient is also discussed.
 This work provides a feasible solution for the design of future field-free spintronic devices.

\bibliographystyle{IEEEtran}
\bibliography{Bibliography.bib}

\end{document}